\documentclass[journal,twocolumn,twoside]{IEEEtran}

\usepackage{epsfig,scalefnt,multirow}
\usepackage{url}
\usepackage{ifthen}
\usepackage{mathtools}
\usepackage{cite}
\usepackage{graphicx}
\usepackage{amssymb}
\usepackage{caption}
\usepackage{booktabs}
\usepackage{amsmath}
\usepackage{epstopdf}
\usepackage{algorithm}
\usepackage{algpseudocode}
\usepackage{cases}
\usepackage{xcolor}
\usepackage{multicol}
\usepackage{stfloats}
\usepackage{blindtext}

  \def\cC{{\mathcal{C}}} 
   \def\cH{{\mathcal{H}}}
   
 \def\cN{{\mathcal{N}}}  
   \def\cT{{\mathcal{T}}}

\def\argmin{\mathop{\mathrm{argmin}}}

\def\b0{{\pmb{0}}} 

 \def\bff{{\mathbf{f}}}  \def\bh{{\mathbf{h}}}
   
   \def\bp{{\mathbf{p}}}
\def\bq{{\mathbf{q}}}   
  \def\bw{{\mathbf{w}}} \def\bx{{\mathbf{x}}}
\def\by{{\mathbf{y}}} \def\bz{{\mathbf{z}}}  

\def\bA{{\mathbf{A}}}   
 \def\bF{{\mathbf{F}}}  \def\bH{{\mathbf{H}}}
\def\bI{{\mathbf{I}}}   
   
 \def\bR{{\mathbf{R}}}  
   
 \def\bZ{{\mathbf{Z}}}

\DeclarePairedDelimiter\norm{\lVert}{\rVert}

\begin{document}
	\title{Massive MIMO Channel Prediction \\ Via Meta-Learning and Deep Denoising:\\ Is a Small Dataset Enough?}
	
	\author{Hwanjin~Kim, \IEEEmembership{Member,~IEEE}, Junil~Choi, \IEEEmembership{Senior Member,~IEEE}, \\and David~J.~Love, \IEEEmembership{Fellow,~IEEE}
	\thanks{This research was partly supported by Samsung Electronics Co., Ltd. (IO210204-08389-01), by Basic Science Research Program through the National Research Foundation of Korea (NRF) funded by the Ministry of Education (2021R1A6A3A13045274), and by Institute of Information \& communications Technology Planning \& Evaluation (IITP) grant funded by the Korea government (MSIT) (No.2020-0-01882, The Development of Dangerous status recognition Platform in building based on WiFi Low Power Wireless signal sensing without sensor or video camera).}
	\thanks{H. Kim and J. Choi are with the School of Electrical Engineering, Korea Advanced Institute of Science and Technology, Daejeon 34141, South Korea (e-mail: jin0903@kaist.ac.kr; junil@kaist.ac.kr).}
	\thanks{D. J. Love is with the School of Electrical and Computer Engineering, Purdue University (e-mail: djlove@purdue.edu).}
	}

\maketitle

\begin{abstract}
	Accurate channel knowledge is critical in massive multiple-input multiple-output (MIMO), which motivates the use of channel prediction. Machine learning techniques for channel prediction hold much promise, but current schemes are limited in their ability to adapt to changes in the environment because they require large training overheads. To accurately predict wireless channels for new environments with reduced training overhead, we propose a fast adaptive channel prediction technique based on a meta-learning algorithm for massive MIMO communications. We exploit the model-agnostic meta-learning (MAML) algorithm to achieve quick adaptation with a small amount of labeled data. Also, to improve the prediction accuracy, we adopt the denoising process for the training data by using deep image prior (DIP). Numerical results show that the proposed MAML-based channel predictor can improve the prediction accuracy with only a few fine-tuning samples. The DIP-based denoising process gives an additional gain in channel prediction, especially in low signal-to-noise ratio regimes.
\end{abstract}


\section{Introduction}
\IEEEPARstart{M}{assive} multiple-input multiple-output (MIMO) systems are expected to be critical to nearly all future broadband wireless systems because of the ever-growing demand for increased network spectral efficiency \cite{Marzetta06}. Accurate channel knowledge at the base station (BS) is often critical to maximizing massive MIMO performance. This is problematic because user equipment (UE) mobility can cause the BS's channel state information (CSI) to become quickly outdated \cite{Papa17, Truong13}. One possible solution is to predict the current channel with the past CSI \cite{Choi14, Kong15, Larew19, Chou21}.

In 3GPP Release 18, a new study item on artificial intelligence (AI)/machine learning (ML) for the new radio (NR) air interface has been started to investigate the benefit of AI/ML on wireless communication systems \cite{AIML}. Utilizing advances in AI/ML, ML-based channel estimators and predictors have recently been proposed for massive MIMO communications in \cite{Dong19, Jiang20, Yuan20, Bogale20, Kim21, Wu21}. By exploiting the frequency and spatial correlations, a wideband channel estimator based on a deep convolutional neural network (CNN) was proposed in \cite{Dong19}. In \cite{Jiang20}, a recurrent neural network (RNN)-based long-range MIMO channel predictor was developed. Also, a CNN combined with autoregressive (AR) predictor and a CNN-based predictor using an RNN-based input were proposed in \cite{Yuan20}. For vehicular-to-infrastructure (V2I) networks, adaptive channel prediction, beamforming, and scheduling were proposed in \cite{Bogale20}. A multi-layer perceptron (MLP)-based channel prediction via the estimated mobility of UE was developed in \cite{Kim21}. To leverage the temporal correlation and the sparse structure of channels, a complex-valued neural network (CVNN)-based predictor was developed for orthogonal frequency division multiplexing (OFDM) communication systems \cite{Wu21}. These predictors, however, require high training overhead to obtain accurate CSI prediction results. Furthermore, a well-trained neural network (NN) model could suffer from significant performance degradation when test environments are different from the training environment.  It may be possible to mitigate these issues by exploiting more advanced ML techniques.

For adaptive ML-based techniques, meta-learning-based schemes have been developed in \cite{Andrychowicz16, Ravi17}. The basic principle of meta-learning is \textit{learning to learn} since it aims to train how to learn a network. The meta-learning algorithm makes it possible to adapt to a new environment quickly without training an NN from scratch. With this meta-learning adaptation, it may be possible to predict the current wireless channel of a new environment by using only a few training samples.

So far, meta-learning algorithms have been widely used for symbol detection, channel estimation, and beamforming adaptation in \cite{HWu19, Mao19, Zhang21, Park21, Yuan21, Xia21, Long21, Zhang22, Yang20, Zeng21}. In \cite{HWu19}, a meta-learning-based two-layer sensing and learning algorithm was proposed for adaptive field sensing and reconstruction. Robust channel estimation using meta neural networks (RoemNet) was implemented to estimate the CSI of OFDM systems in \cite{Mao19}, and an online training mechanism was developed for the long short-term memory (LSTM) optimizer based on meta-learning in \cite{Zhang21}. Offline and online meta-learning frameworks for Internet-of-Things (IoT) systems were proposed in \cite{Park21}. Fast beamforming techniques using meta-learning were considered in \cite{Yuan21, Xia21, Long21, Zhang22}, and downlink channel prediction using uplink CSI for frequency-division duplexing (FDD) MIMO systems exploiting meta-learning was proposed in \cite{Yang20, Zeng21}. In this paper, we adopt the model-agnostic meta-learning (MAML) algorithm \cite{Finn17}, which is the optimization-based meta-learning approach, for fast adaptive channel prediction in massive MIMO communications. In the MAML algorithm, we divide the UEs into two sets, which are a training set and a testing set. To be specific, our proposed channel predictor is first trained with the measurement data from the training set. Then, the trained predictor can adaptively predict the channels of the testing set (which are different from the UEs used for the training) with only a small number of measurement data. The MAML algorithm is suitable to our problem of interest since it learns an initialization of model parameters, allowing quick adaptation to a new environment using a small amount of data.  

To improve the performance of ML-based techniques, the data denoising process is crucial since the noise-corrupted data in the training phase may cause performance degradation \cite{Kim20}. The least square (LS) or minimum mean-squared error (MMSE) processes are typically used for denoising the training data \cite{Jin19, Soltani19}. However, the LS-based denoising process has limited performance, and the MMSE-based denoising process needs prior knowledge of channel statistics, e.g., channel covariance \cite{Balevi20}. Different from the conventional denoising process, ML-based denoising processes have been proposed in \cite{Zhang17, He18, Zhang20, Ye20}. However, the ML-based denoising approaches have high computational complexity due to the deep neural network (DNN)-based architecture using a large number of labeled data \cite{He18, Zhang20, Ye20}. Moreover, these ML-based approaches use the true channel for the training phase, which is impractical. Different from these works, we adopt the deep image prior (DIP) to resolve these problems \cite{Ulyanov18}. The DIP-based denoising process neither requires the channel statistics nor the true channel data; instead, it only uses the measurement data, i.e., noise-corrupted data, for updating the model, which is well suited to wireless communication environments.

In this paper, we show that it is possible to formulate the massive MIMO channel prediction problem as an optimization problem by exploiting the previous measurements. We then propose a fast adaptive channel predictor based on meta-learning for massive MIMO. We adopt the MAML algorithm for the meta-learning since its structure well incorporates out-of-distribution data. Using the temporally correlated measurement data, the channel prediction using the MAML algorithm consists of three stages: meta-training, meta-adaptation, and meta-testing stages. In the meta-training stage, the meta-learner aims to optimize global network parameters. In the meta-adaptation stage, the network parameters are refined using only a few adaptation samples from new environments where the new environments refer to the UEs not considered during the meta-training stage. With these fine-tuned network parameters, the BS predicts the channel of these new UEs in the meta-testing stage. To obtain better prediction results, especially in low signal-to-noise ratio (SNR) regimes, we also exploit the CNN architecture-based DIP to denoise the training data. The numerical results reveal that the proposed MAML-based channel predictor outperforms the conventional ML-based predictor. The DIP-based denoising process can give further improvements in channel prediction performance.

The remainder of the paper is structured as follows. We describe a system model and an optimization problem for the channel prediction in Section \ref{Section:system model}. We propose the MAML-based predictor in Section \ref{Section:MAML} and explain the DIP-based denoising process in Section \ref{Section:DIP}. In Section \ref{Section:numerical result}, we examine the computational complexity of the channel predictors and present numerical results to validate our algorithms and analysis. Finally, concluding remarks are provided in Section \ref{Section:conclusion}.

\textbf{Notation:} Upper case and lower case boldface letters indicate matrices and column vectors, respectively. The transpose, conjugate transpose, and inverse of matrix $\bA$ are represented by $\bA^{\mathrm{T}}$, $\bA^{\mathrm{H}}$, and $\bA^{-1}$, respectively. ${\boldsymbol{0}}_{m}$ denotes the $m \times 1$ all zero vector, and $\bI_m$ is used for the $m \times m$ identity matrix. $\cC \cN(\bx,\bR)$ represents the complex Gaussian distribution having mean $\bx$ and covariance $\bR$. The set of all $m \times n$ real matrices is represented by ${\mathbb{R}}^{m \times n}$. $\norm{\cdot}$ denotes the $\ell_2$-norm of vector, and $|{\cdot}|$ represents the amplitude of scalar. $\lfloor x \rfloor$ denotes the floor function of $x$. $\mathcal{O}(\cdot)$ represents the Big-O notation. $\mathbb{E}[\cdot]$ denotes the expectation.

\begin{figure}[tbp]
	\centering
	\includegraphics[width=8 cm]{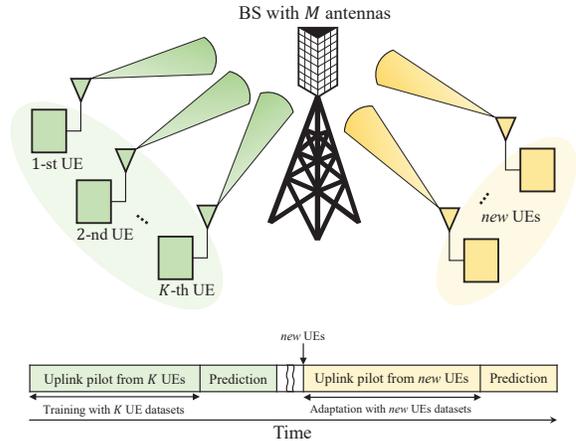}
	\caption{Massive MIMO system consisting of a BS with $M$ antennas and $K$ UEs with a single antenna each. The BS trains the NN with the pilot signals from $K$ UEs, then predicts the channels of \textit{new} UEs with adaptation data from a small amount of pilot signals from the \textit{new} UEs.}
	\label{figure1}
\end{figure}

\section{System Model and Problem Formulation}\label{Section:system model}

\subsection{System Model and Considering Scenario}
In Fig. \ref{figure1}, we consider a single-cell uplink massive MIMO system consisting of a BS with $M$ antennas and $K$ UEs with a single-antenna each. We assume that the BS trains the NN with the uplink pilot signals from $K$ UEs. Then, the BS aims to predict the channels of \textit{new} UEs based on the pre-trained NN and a few adaptation samples from the \textit{new} UEs pilot signals. Note that we define the \textit{new} UEs as the UEs that are served by the BS for the first time. Since the BS predicts each UE channel separately, we consider only the $k$-th UE's input-output expression
\begin{align}
\by_{n,k}=\sqrt{\rho}\bh_{n,k}x_{n,k}+\bw_{n,k}, \label{measurement}
\end{align} where $\rho$ is the SNR, $\bh_{n,k}$ is the channel between the BS and the $k$-th UE, $x_{n,k}$ is the pilot signal, and the complex Gaussian noise is denoted as $\bw_{n,k}\sim \mathcal{C}\mathcal{N}(\boldsymbol{0}_{M},\bI_{M})$. 
Also, the received signal from the $k_\text{T}$-th \textit{new} UE at the BS during the $n$-th time slot is expressed as
\begin{align}
\by_{n,k_\text{T}}=\sqrt{\rho}\bh_{n,k_\text{T}}x_{n,k_\text{T}}+\bw_{n,k_\text{T}}, \label{measurement2}
\end{align} where $k_\text{T}\in \mathcal{K}_{\text{new}}$ is the \textit{new} UE index, and $\mathcal{K}_{\text{new}}$ is the index set of \textit{new} UEs.

\subsection{Problem Formulation}

\begin{figure*}[tbp]
	\centering
	\includegraphics[width=12 cm]{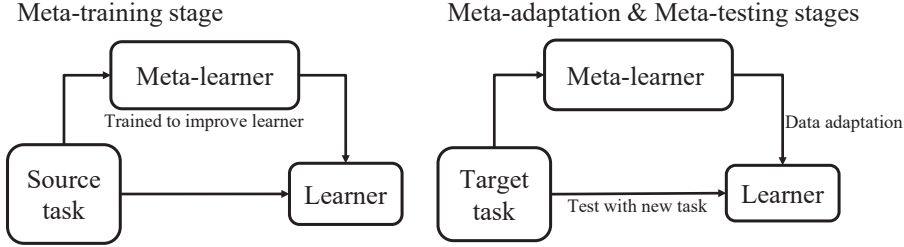}
	\caption{MAML structure: meta-training, meta-adaptation, and meta-testing stages.}
	\label{figure2}
\end{figure*}

To predict the \textit{new} UE channel $\bh_{n,k_\text{T}}$, we use the temporal correlation of the channels, i.e., based on the proper complexity order of $n_o$, the BS predicts the channel by exploiting the previous $n_o$ measurements. The optimization problem for the channel prediction is defined as
\begin{alignat}{3}
&\text{minimize} &&~{\left\|{\bh}_{n+1,k_\text{T}}-\hat{{\bh}}_{n+1,k_\text{T}}\right\|}^2 \label{optimization}\\
&\text{subject to}&&~{\hat{{\bh}}}_{n+1,k_\text{T}}=f\big({\by}_{n-n_o+1,k_\text{T}},\cdots,{\by}_{n,k_\text{T}}\big), \notag
\end{alignat} 
where ${\hat{\bh}}_{n+1,k_\text{T}}$ is the predicted channel for $k_\text{T}$-th UE at the $(n+1)$-th time slot produced by the prediction function $f(\cdot)$. Note that using the true channel $\bh_{n+1,k_\text{T}}$ as the target value for the optimization problem in (\ref{optimization}) is impractical. Therefore, we assume that the BS only exploits realistic measurement data for the target value as 
\begin{alignat}{3}
&\text{minimize} &&~{\left\|\bh_{n+1,k_\text{T}}^\text{LS}-\hat{{\bh}}_{n+1,k_\text{T}}\right\|}^2 \label{optimization2}\\
&\text{subject to} &&~{\hat{{\bh}}}_{n+1,k_\text{T}}=f\left({\bh}_{n-n_o+1,k_\text{T}}^\text{LS},\cdots,{\bh}_{n,k_\text{T}}^\text{LS}\right), \notag
\end{alignat}
where $\bh_{n,k}^\text{LS}$ is the least square (LS) channel estimate given by
\begin{align}
\bh_{n,k}^\text{LS}=\frac{1}{\sqrt{\rho}x_{n,k}}{\by}_{n,k}=\bh_{n,k}+\bw_{n,k}',\quad\forall n,\forall k, \label{LS estimate}
\end{align} with $\bw_{n,k}'=\frac{1}{\sqrt{\rho}x_{n,k}}\bw_{n,k}$. We assume that the SNR $\rho$ is a long-term statistic and can be perfectly estimated at the BS \cite{Hung10}. From the NN training perspective, the loss function can be defined as the sum of mean-squared error (MSE) between the LS channel estimate and predicted channel,  
\begin{align}
\text{Loss}=\frac{1}{N}\sum_{n=1}^N \left\|\bh_{n+1,k}^\text{LS}-\hat\bh_{n+1,k}\right\|^2, \label{loss function}
\end{align}
where $N$ denotes the number of samples. The loss function in (\ref{loss function}) will be used for the MAML algorithm in Section \ref{Section:MAML}. In the following sections, we will use the terms $\textit{received signals}$ and $\textit{measurements}$ interchangeably.

\section{MAML-Based Channel Prediction}\label{Section:MAML}

To obtain accurate channel prediction in (\ref{optimization}) using conventional ML techniques for various scenarios, e.g., different UE configurations, the BS requires a large amount of training overhead for each scenario \cite{Dong19, Jiang20, Yuan20}. It is crucial to resolve this training issue for ML techniques to work in practice, and we exploit the meta-learning algorithm to address this problem. With the meta-learning algorithm, the BS can predict the channels of various UE configurations more quickly using a small number of adaptation samples. 

\subsection{MAML Structure and Task}\label{meta learning task}
Among many possible meta-learning algorithms, we adopt the MAML algorithm proposed in \cite{Finn17} that is used in various neural networks. The MAML algorithm has a hierarchical structure with the meta-learner and the learner, consisting of three stages: 1) meta-training stage, 2) meta-adaptation stage, and 3) meta-testing stage as in Fig. \ref{figure2}. Following the terminologies of meta-learning, we define a meta-learning task $\mathcal{T}$, which consists of a dataset and a loss function $\mathcal{T}=\{\mathbb{D},\text{Loss}_\mathbb{D}\}$ \cite{Timothy21}, as the prediction of a target UE channel exploiting previous measurements. The meta-learning task $\mathcal{T}$ is also composed of a source task $\mathcal{T}_\text{S}$ for the meta-training stage and a target task $\mathcal{T}_\text{T}$ for the meta-adaptation and meta-testing stages. We will define $\mathbb{D}$ and $\text{Loss}_\mathbb{D}$ of the task $\mathcal{T}$ and the relation among $\mathcal{T}_\text{S}$, $\mathcal{T}_\text{T}$, $\mathbb{D}$, and $\text{Loss}_\mathbb{D}$ in detail in Sections \ref{meta dataset} and \ref{Meta-training stage}.

The BS first trains the meta-learner with the source task. Then, the meta-learner helps the learner adjust to a new task utilizing only a small number of adaptation samples from the target task. The meta-learner aims to learn the inductive bias while the learner adapts to a new task with this inductive bias. 

\subsection{Definition of MAML Datasets}\label{meta dataset}
For each stage of the MAML algorithm, we use an independent dataset. We define the LS channel estimates from some UEs in the meta-training stage as the source dataset $\mathbb{D}_\text{S}$ in $\mathcal{T}_\text{S}$, and the LS channel estimates from other UEs (that are different from the UEs used during the meta-training stage) in the meta-adaptation and meta-testing stages as the target dataset $\mathbb{D}_\text{T}$ in $\cT_\text{T}$. We refer to the support set as $\mathbb{D}_{\text{Sup}}$ for the training data and the query set as $\mathbb{D}_{\text{Que}}$ for the validation data during the meta-training stage. To prevent the network model from overfitting, we split the support set $\mathbb{D}_{\text{Sup}}$ and the query set $\mathbb{D}_{\text{Que}}$, i.e., $\mathbb{D}_{\text{Sup}} \cap \mathbb{D}_{\text{Que}}=\emptyset$. We define the datasets for the meta-adaptation and meta-testing stages as $\mathbb{D}_{\text{Ad}}$ and $\mathbb{D}_{\text{Te}}$, respectively. Also, we assume that no sample in $\mathbb{D}_{\text{Te}}$ appears in $\mathbb{D}_{\text{Ad}}$, i.e., $\mathbb{D}_{\text{Te}} \cap \mathbb{D}_{\text{Ad}}=\emptyset$. Then, it is clear that $\mathbb{D}_\text{S}=\mathbb{D}_{\text{Sup}} \cup \mathbb{D}_{\text{Que}}$ and $\mathbb{D}_\text{T}=\mathbb{D}_{\text{Ad}} \cup \mathbb{D}_{\text{Te}}$. Note that the distribution in the target dataset $\mathbb{D}_{\text{T}}$ is different from the distribution in the source dataset $\mathbb{D}_\text{S}$. Thus, all MAML datasets are non-overlapping.

\begin{figure*}[t]
	\centering
	\includegraphics[width=16 cm]{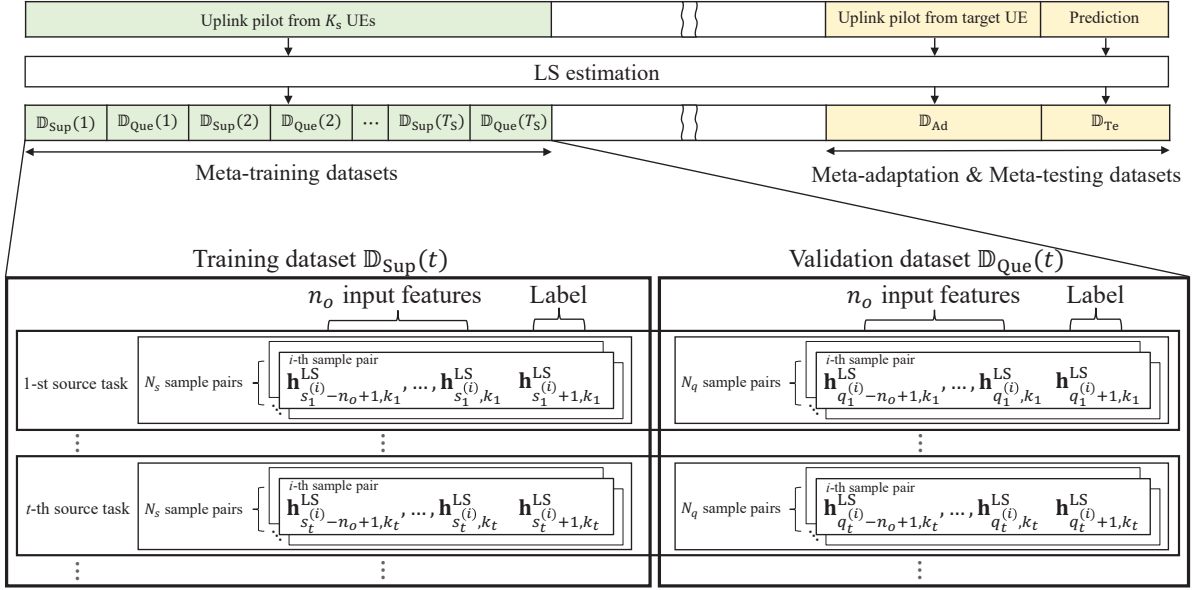}
	\caption{MAML datasets consist of the meta-training, meta-adaptation, and meta-testing datasets. After the LS estimation process of the uplink pilot signals, the BS allocates the LS channel estimates into each dataset. The meta-training datasets include the training and validation datasets. Each sample pair in the task consists of $n_o$ input features and one label to exploit the temporal correlation of channels.}
	\label{figure3}
\end{figure*}

Fig. \ref{figure3} reveals the MAML datasets, which consist of the meta-training, meta-adaptation, and meta-testing datasets. The BS collects the uplink pilot signals from multiple UEs, then performs the LS estimation as in (\ref{LS estimate}). Finally, the LS channel estimates are binned into each dataset. In the meta-training stage, the BS uses the total number of $T_\text{S}=T_uK_s$ source tasks $\{\cT_\text{S}(t)\}_{t=1}^{T_\text{S}}$, where $T_u$ is the number of source tasks per UE, and $K_s$ is the number of UEs for the source task. Each dataset of the $t$-th source task $\mathbb{D}_{\text{S}}(t)$ consists of two disjoint datasets: the support set $\mathbb{D}_{\text{Sup}}(t)$ and the query set $\mathbb{D}_{\text{Que}}(t)$, i.e., $\mathbb{D}_{\text{S}}(t)=\left\{\mathbb{D}_\text{Sup}(t), \mathbb{D}_\text{Que}(t)\right\}$. We denote the support set of $t$-th source task, which includes $N_s$ labeled data, as $\mathbb{D}_{\text{Sup}}(t)=\left\{\left\{\bp_{\text{Sup},t}^{(i)}, \bq_{\text{Sup},t}^{(i)}\right\}\right\}_{i=1}^{N_s}$, where $\left\{\bp_{\text{Sup},t}^{(i)}, \bq_{\text{Sup},t}^{(i)}\right\}$ is the $i$-th sample pair in the support set. We use $n_o$ input features $\bp_{\text{Sup},t}^{(i)}=\left\{\bh_{s^{(i)}_t-n_o+1,k_t}^\text{LS},...,\bh_{s^{(i)}_t,k_t}^\text{LS}\right\}$ and one label $\bq_{\text{Sup},t}^{(i)}=\bh_{s^{(i)}_t+1,k_t}^\text{LS}$, where $s^{(i)}_t$ is the $i$-th sample index of the support set, and $k_t=\lfloor \frac{t-1}{T_u} \rfloor+1$ is the UE index for the $t$-th source task.

Similarly, the query set of the $t$-th source task with $N_q$ labeled data is denoted as $\mathbb{D}_{\text{Que}}(t)=\left\{\left\{\bp_{\text{Que},t}^{(i)}, \bq_{\text{Que},t}^{(i)}\right\}\right\}_{i=1}^{N_q}$, where $\left\{\bp_{\text{Que},t}^{(i)}, \bq_{\text{Que},t}^{(i)}\right\}$ is the $i$-th samples pair in the query set. Also, each sample pair includes $n_o$ input features $\bp_{\text{Que},t}^{(i)}=\left\{\bh_{q^{(i)}_t-n_o+1,k_t}^\text{LS},...,\bh_{q^{(i)}_t,k_t}^\text{LS}\right\}$ and the corresponding label $\bq_{\text{Que},t}^{(i)}=\bh_{q^{(i)}_t+1,k_t}^\text{LS}$, where $q^{(i)}_t$ is the $i$-th sample index of the query set for the $t$-th source task. 

In the target task $\cT_\text{T}$, we define the meta-adaptation dataset with the number of adaptation samples $N_{\text{ad}}$ as $\mathbb{D}_\text{Ad}=\left\{\left\{\bp_{\text{Ad}}^{(i)}, \bq_{\text{Ad}}^{(i)}\right\}\right\}_{i=1}^{N_\text{ad}}$, where $\bp_{\text{Ad}}^{(i)}=\left\{\bh_{a^{(i)}_\text{T}-n_o+1,k_\text{T}}^\text{LS},...,\bh_{a^{(i)}_\text{T},k_\text{T}}^\text{LS}\right\}$ and $\bq_{\text{Ad}}^{(i)}=\bh_{a^{(i)}_\text{T}+1,k_\text{T}}^\text{LS}$. Note that $a^{(i)}_\text{T}$ is the $i$-th target sample index of the meta-adaptation dataset, and $k_\text{T}$ is the target UE index of the meta-adaptation dataset, which is the same as the \textit{new} UE index in (\ref{measurement2}). We also define the meta-testing dataset with the number of test samples $N_{\text{te}}$ as $\mathbb{D}_\text{Te}=\left\{\left\{\bp_{\text{Te}}^{(i)}, \bq_{\text{Te}}^{(i)}\right\}\right\}_{i=1}^{N_\text{te}}$, where $\bp_{\text{Te}}^{(i)}=\left\{\bh_{b^{(i)}_\text{T}-n_o+1,k_\text{T}}^\text{LS},...,\bh_{b^{(i)}_\text{T},k_\text{T}}^\text{LS}\right\}$ and $\bq_{\text{Te}}^{(i)}=\bh_{b^{(i)}_\text{T}+1,k_\text{T}}$ with the $i$-th target sample index of the meta-testing dataset $b^{(i)}_\text{T}$.

\subsection{Meta-Training Stage} \label{Meta-training stage}
The objective of a meta-learner is to acquire the inductive bias from the entire source tasks $\{\cT_\text{S}(t)\}_{t=1}^{T_\text{S}}$ for fast adaptation in the meta-training stage. The meta-learner parameters are updated using inner-task and outer-task update processes. In the inner-task update, the BS trains the NN parameters of each task in the corresponding batch, where the batch is the group of source tasks for efficiently updating gradient steps. The BS groups the source tasks by the batch size of $V$ and updates the NN parameters with $V$ source tasks in each iteration. The BS uses the mini-batch stochastic gradient descent (SGD) method \cite{Goyal17} using the batch size of $V$ to update the inner-task parameters of the $t$-th source task, $\boldsymbol{\Omega}_{\text{Tr},t}$, 
\begin{align}
\boldsymbol{\Omega}_{\text{Tr},t}\leftarrow \boldsymbol{\Omega}_{\text{Tr},t}-\alpha\nabla_{\boldsymbol{\Omega}_{\text{Tr},t}}\text{Loss}_{\mathbb{D}_{\text{Sup}}(t)}(\boldsymbol{\Omega}_{\text{Tr},t}),~t=1,...,V, \label{inner-task}
\end{align}where $\alpha$ represents the inner-task learning rate and $\text{Loss}_{\mathbb{D}_{\text{Sup}}(t)}$ denotes the loss function on $\mathbb{D}_{\text{Sup}}(t)$. We use the MSE between the target value $\bq_{\text{Sup},t}^{(i)}$ and the predicted value $\hat{\bq}_{\text{Sup},t}^{(i)}$ as the loss function
\begin{align}
\text{Loss}_{\mathbb{D}_{\text{Sup}}(t)}=\frac{1}{N_s}\sum_{i=1}^{N_s} \left\|{\bq}_{\text{Sup},t}^{(i)}-\hat{\bq}_{\text{Sup},t}^{(i)}\right\|^2. \label{loss}
\end{align} In (\ref{loss}), the BS uses the LS channel estimate $\bq_{\text{Sup},t}^{(i)}=\bh_{s^{(i)}_t+1,k_t}^\text{LS}$ as the target value. Note that $\bh_{s^{(i)}_t+1,k_t}^\text{LS}$ is corrupted with the noise, and we exploit the DIP architecture to denoise the LS channel estimate in Section \ref{Section:DIP}.

After the inner-task update, the outer-task update is performed to optimize the global network parameters $\boldsymbol{\Omega}$. In the outer-task update, the BS updates the global network parameters $\boldsymbol{\Omega}$ to minimize the sum of the loss functions of tasks on $\mathbb{D}_{\text{Que}}(t)$, i.e.,
\begin{align}
\sum_{t=1}^V\text{Loss}_{\mathbb{D}_{\text{Que}}(t)}(\boldsymbol{\Omega}_{\text{Tr},t}), \label{outer-task}
\end{align} where $\text{Loss}_{\mathbb{D}_{\text{Que}}(t)}$ is the loss function on the query set $\mathbb{D}_{\text{Que}}(t)$ as in (\ref{loss}). The global network parameters $\boldsymbol{\Omega}$ is updated by the adaptive moment estimation (ADAM) optimizer \cite{Kingma15} with the outer-task learning rate $\beta$. 

The BS performs the inner-task and outer-task updates iteratively according to the number of epochs $N_\text{epoch}$, which indicates the total number of passes through the entire training dataset. Thus, the total number of iterations for the meta-training stage is $N_\text{epoch}T_\text{S}/V$ since the number of iterations in each epoch is $T_\text{S}/V$. With these definitions, now we can concretely define the source task $\cT_\text{S}=\{\mathbb{D}_{\text{Sup}}, \mathbb{D}_{\text{Que}}, \text{Loss}_{\mathbb{D}_{\text{Sup}}}, \text{Loss}_{\mathbb{D}_{\text{Que}}}\}$.

\begin{algorithm}[tbp]
	\renewcommand{\arraystretch}{1.5}
	\newcommand{\algrule}[1][.2pt]{\par\vskip.5\baselineskip\hrule height #1\par\vskip.5\baselineskip}
	\begin{algorithmic}[1]
		\caption{MAML-Based Channel Predictor}
		\State $\mathbf{Input}$: Source task $\{\cT_\text{S}(t)\}_{t=1}^{T_\text{S}}$, Target task $\cT_{\text{T}}$, inner-task learning rate $\alpha$, outer-task learning rate $\beta$, batch size $V$, number of epochs $N_\text{epoch}$
		\State $\mathbf{Output}$: Predicted channel
		\algrule
		\State ${\textit{Meta-training stage}}$:\\
		Randomly initialize the neural network parameters
		\For{$j=1,...,N_\text{epoch}T_\text{S}/V$} {}
		\State Randomly sample $V$ batch of tasks from $\{\cT_\text{S}(t)\}_{t=1}^{T_\text{S}}$
		\State Generate datasets $\{\mathbb{D}_{\text{Sup}}(t)\}_{t=1}^V$ and $\{\mathbb{D}_{\text{Que}}(t)\}_{t=1}^V$
		\For{$t=1,...,V$} {}
		\State Update $\boldsymbol{\Omega}_{\text{Tr},t}$ by (\ref{inner-task})  with $\mathbb{D}_{\text{Sup}}(t)$
		\EndFor
		\State Update $\boldsymbol{\Omega}$ to minimize (\ref{outer-task})
		\EndFor
		\algrule
		\State $\textit{Meta-adaptation stage}$:
		\State Generate datasets $\mathbb{D}_\text{Ad}$ and $\mathbb{D}_\text{Te}$ from $\cT_\text{T}$ \\
		Load the meta-trained network parameters
		\For{$j=1,...,T_{\text{ad}}$} {}
		\State Update $\boldsymbol{\Omega}_{\text{Ad}}$ by (\ref{adaptation}) with  $\mathbb{D}_{\text{Ad}}$
		\EndFor
		\algrule
		\State $\textit{Meta-testing stage}$:\\
		Predict the channel based on $\mathbb{D}_{\text{Te}}$ and $\boldsymbol{\Omega}_{\text{Ad}}$
	\end{algorithmic}
\end{algorithm} 

\subsection{Meta-Adaptation and Meta-Testing Stages}
In the meta-adaptation stage, the BS updates the network parameters to adapt to a new task quickly using the adaptation dataset $\mathbb{D}_{\text{Ad}}$ based on the pre-trained global network parameters $\boldsymbol{\Omega}$. The adaptation parameters $\boldsymbol{\Omega}_{\text{Ad}}$ are updated by the SGD method with the number of adaptation samples $N_\text{ad}$ as
\begin{align}
\boldsymbol{\Omega}_{\text{Ad}}\leftarrow \boldsymbol{\Omega}_{\text{Ad}}-\alpha\nabla_{\boldsymbol{\Omega}_{\text{Ad}}}\text{Loss}_{\mathbb{D}_{\text{Ad}}}(\boldsymbol{\Omega}_{\text{Ad}}), \label{adaptation}
\end{align} where  $\text{Loss}_{\mathbb{D}_{\text{Ad}}}$ is the loss function of adaptation dataset $\mathbb{D}_{\text{Ad}}$. After finishing the fine-tuning with $T_\text{ad}$ gradient steps, the meta-testing stage gives the predicted channel using $\boldsymbol{\Omega}_{\text{Ad}}$ and $\mathbb{D}_{\text{Te}}$.
The proposed MAML-based channel prediction algorithm is summarized in Algorithm 1. We can also rigorously define the target task $\cT_\text{T}$ with the terminologies used in this subsection as $\cT_\text{T}=\{\mathbb{D}_{\text{Ad}}, \mathbb{D}_{\text{Te}}, \text{Loss}_{\mathbb{D}_{\text{Ad}}}, \text{Loss}_{\mathbb{D}_{\text{Te}}}\}$.

\begin{figure*}[t]
	\centering
	\includegraphics[width=16 cm]{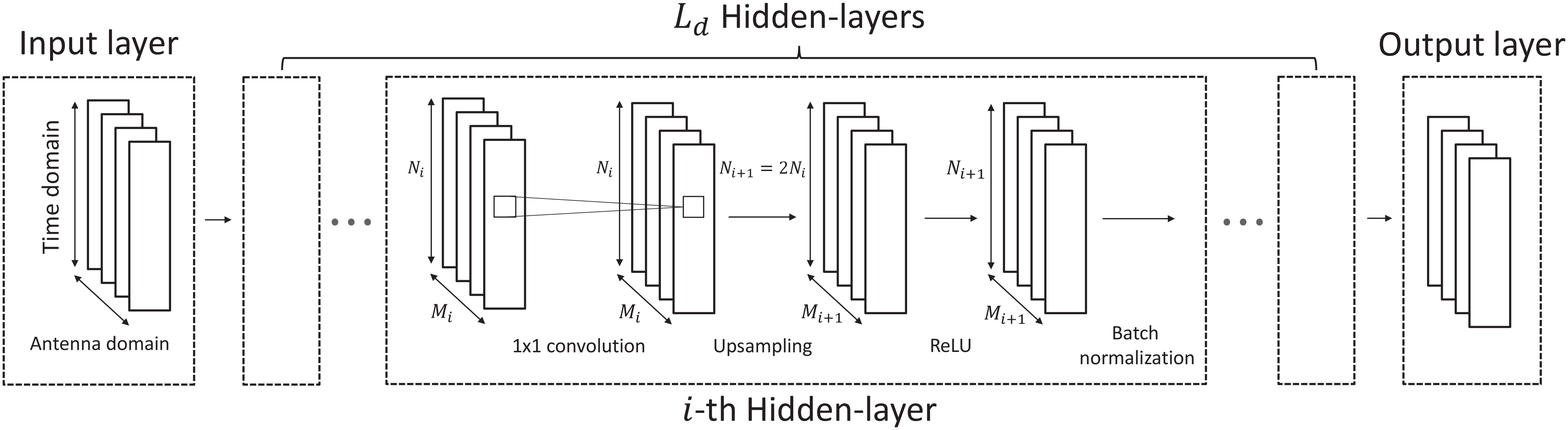}%
	\caption{DIP-based denoising process architecture.}
	\label{figure4}
\end{figure*}

\section{DIP-Based Denoising Process}\label{Section:DIP}
In practice, the received signal contains Gaussian noise, i.e., $\by_{n,k}=\sqrt{\rho}\bh_{n,k}x_{n,k}+\bw_{n,k}$. Thus, before exploiting this input-output relation to collect training data, we need to apply the denoising process to achieve better prediction results. 

The denoising process for the noise-corrupted signal $\bh_{n,k}^\text{LS}$ as in (\ref{LS estimate}) is a kind of \textit{inverse problem} \cite{Arridge19, He21}. With the prior of the channel statistics, we can get a clean signal by maximizing a likelihood function. However, the prior of the channel statistics is hard to obtain in practice. Thus, we solve the inverse problem by minimizing the MSE with the DIP architecture \cite{Ulyanov18} 
\begin{align}
	&\boldsymbol{\Phi}^*=\argmin_{\boldsymbol{\Phi}}\norm{\bh_{n,k}^\text{LS}-\hat{\bh}_{n,k}^\text{LS}}^2\label{DIP_mini}\\
	&\text{subject to} ~\hat{\bh}_{n,k}^\text{LS}=g_{\boldsymbol{\Phi}}(\bz)\notag,
\end{align} where $g_{\boldsymbol{\Phi}}(\bz)$ is the NN function with the network parameters $\boldsymbol{\Phi}$ and the input $\bz$. The underline idea of DIP is that the NN model is better suited to the structured signal than the random noise. Thus, the DIP architecture plays a role as the prior of the denoising process. The minimization problem in (\ref{DIP_mini}) does not need to have any statistical knowledge, and the solution is obtained by using the gradient descent on the NN function without any training in advance. Thus, we can obtain the denoised data with low-complexity using the untrained network.

For the DIP-based denoising process, we stack the LS channel estimates in the time domain as follows. The LS channel estimates in (\ref{LS estimate}) are reformulated as the $2$-dimensional data $\bH^\text{LS}$,
\begin{align}
\bH^\text{LS}=[[\bH_{2\text{D}}^\text{LS}[m,n]]_{m=1}^M]_{n=1}^N,
\end{align} where $\bH_{2\text{D}}^\text{LS}[m,n]$ is the LS channel estimate at the $m$-th BS antenna during the $n$-th time slot. Since the DIP architecture only supports real-valued data, the real and imaginary components of the 2-dimensional data are stacked into the BS antenna domain. This reformulated form of $\bH^\text{LS}$ is defined as $\boldsymbol{\cH}^\text{LS}\in \mathbb{R}^{2M\times N}$. 

In Fig. \ref{figure4}, the DIP architecture contains an input-layer, $L_d$ hidden-layers, and an output-layer. Also, each hidden-layer has four components, which are the $1\times 1$ convolutional layer, upsampling layer, rectified linear unit (ReLU) activation layer, and batch normalization layer. The $i$-th hidden layer for $1\leq i \leq L_d-1$ is given as
\begin{align}
g_{\boldsymbol{\phi}_i}={\textbf{Batch}}({\textbf{ReLU}}({\textbf{Upsample}}(\boldsymbol{\phi}_i\circledast \bZ_i))),
\end{align} where $\boldsymbol{\phi}_i$ are the model parameters of $i$-th hidden layer, $\circledast$ denotes the $1\times 1$ convolution operation, and $\bZ_i$ is the input of $i$-th hidden-layer. Note that the dimensions of the BS antenna domain and the time domain for the $i$-th layer are $M_i$ and $N_i$, respectively. Since we use the $1\times1$ convolution to capture the spatial correlation, the number of network parameters in the DIP architecture is decreased. In the upsampling layer, the time dimension is doubled by the bilinear transformation, i.e., $N_{i+1}=2N_{i}$. Since the channel is temporally correlated, the upsampling layer can  leverage the correlation between the adjacent elements in the time domain. For the last hidden-layer, we set the ReLU activation layer followed by the batch normalization layer as
\begin{align}
g_{\boldsymbol{\phi}_{L_d}}={\textbf{Batch}}({\textbf{ReLU}}(\boldsymbol{\phi}_{L_d}\circledast \bZ_{L_d})),
\end{align}to avoid the vanishing gradients problem. The final output-layer is
\begin{align}
g_{\boldsymbol{\phi}_{L_d+1}}=\boldsymbol{\phi}_{L_d+1}\circledast \bZ_{L_d+1}.
\end{align} The optimization problem for the DIP architecture is given by the $\ell_2$-norm
\begin{align}
\boldsymbol{\Phi}^*=\argmin_{\boldsymbol{\Phi}}\left\|\boldsymbol{\cH}^\text{LS}-\hat{\boldsymbol{\cH}}^\text{LS}\right\|^2, \label{DIP opt}
\end{align} where $\boldsymbol{\Phi}=[\boldsymbol{\phi}_1,\cdots,\boldsymbol{\phi}_{L_d+1}]$, and $\hat{\boldsymbol{\cH}}^\text{LS}=g_{\boldsymbol{\Phi}}(\bZ_1)$ is the estimate of $\boldsymbol{\cH}^\text{LS}$. Note that $\bZ_1$ is a random initial value with the dimension of $M_1\times N_1$. The DIP architecture gives the solution for (\ref{DIP opt})  using the ADAM optimizer with the number of iterations $N_\text{iter}$.

\section{Computational Complexity and Numerical Results}\label{Section:numerical result}
The computational complexity of the proposed MAML channel prediction and the denoising process based on the DIP is first analyzed in this section. Then, we evaluate the prediction performance of the MAML-based predictor compared to the conventional method. For the complexity analysis, we exploit the floating-point operations (FLOPs) with the Big-O notation \cite{Hunger05}.

\subsection{Computational Complexity}
The MAML-based channel prediction with the MLP structure has three levels of complexity, i.e., the complexity of the meta-training, meta-adaptation, and meta-testing stages. In the meta-training stage, the complexity using the number of epochs $N_\text{epoch}$, the total number of source tasks $T_\text{S}$, the number of meta-training sample pairs in each task $N_\text{mt}=N_s+N_q$, the complexity order $n_o$, and the number of hidden-layers $L$ using $n_l$ nodes is given by \cite{Mizutani01}
\begin{align}
&C_\text{MAML-train}\notag\\ 
&=\mathcal{O}\left(N_\text{epoch}T_\text{S}N_\text{mt}\left(n_oM n_1+\sum_{l=1}^{L-1}n_l n_{l+1}+n_L M\right)\right)\notag\\
&\stackrel{(a)}{=}\mathcal{O}\big(N_\text{epoch}T_\text{S}N_\text{mt}\big(\gamma n_oM^2+(L-1)\gamma^2M^2+\gamma M^2\big)\big)\notag\\
&=\mathcal{O}\left(N_\text{epoch}T_\text{S}N_\text{mt}\gamma(n_o+(L-1)\gamma+1) M^2\right),
\end{align} where $(a)$ is from $n_l=\gamma M$ for $1\leq l \leq L$. Note that $\gamma$ is a scaling factor, which depends on the number of BS antennas, for the hidden-layer nodes. In the meta-adaptation stage, the complexity with the number of gradient steps $T_\text{ad}$ and the number of adaptation samples $N_\text{ad}$ becomes 
\begin{align}
C_\text{MAML-adaptation}=\mathcal{O}\left(T_\text{ad}N_\text{ad}\gamma(n_o+(L-1)\gamma+1) M^2\right).
\end{align} In addition, the complexity of the meta-testing stage with the number of test samples $N_\text{te}$ is $\mathcal{O}\left(N_\text{te}\gamma(n_o+(L-1)\gamma+1) M^2\right)$. Finally, the total complexity of the MAML-based predictor becomes
\begin{align}
C_\text{MAML}&=\mathcal{O}\big(N_{\text{epoch}}T_\text{S}N_\text{mt}+T_\text{ad}N_\text{ad}+N_\text{te})\notag\\
&\qquad \cdot\gamma(n_o+(L-1)\gamma+1) M^2\big).
\end{align}

\begin{table*}[t!]
	\centering
	\renewcommand{\arraystretch}{1.3}
	\captionsetup{justification=centering}
	\captionsetup{labelsep=newline}
	\caption{Computational complexity of MAML-based channel prediction and DIP-based denoising process} 
	\resizebox{\linewidth}{!}{%
		\begin{tabular}{l l l l}
			\toprule
			\text{Method} & Stage & \text{Complexity} & \text{Total complexity} \\
			\hline
			\multirow{3}{*}{MAML}& \multicolumn{1}{l}{Train} & \multicolumn{1}{l}{$\mathcal{O}\left(N_\text{epoch}T_\text{S}N_\text{mt}\gamma(n_o+(L-1)\gamma+1) M^2\right)$} & \multirow{3}{*}{$\mathcal{O}\left((N_\text{epoch}T_\text{S}N_\text{mt}+T_{\text{ad}}N_\text{ad}+N_\text{te})\gamma(n_o+(L-1)\gamma+1)M^2\right)$} \\ 
			& \multicolumn{1}{l}{Adaptation} & \multicolumn{1}{l}{$\mathcal{O}\left(T_{\text{ad}}N_\text{ad}\gamma(n_o+(L-1)\gamma+1) M^2\right)$} & \\  
			& \multicolumn{1}{l}{Test} & \multicolumn{1}{l}{$\mathcal{O}\left(N_\text{te}\gamma(n_o+(L-1)\gamma+1) M^2\right)$} & \\ \hline
			\multirow{1}{*}{DIP} & \multicolumn{1}{l}{-} &\multicolumn{1}{l}{$\mathcal{O}\left(N_\text{iter}N_tN_f(2N_f+M)\right)$} & \multirow{1}{*}{-} \\ 		
			\bottomrule
	\end{tabular}}
	{\label{table1}}
\end{table*}

The total complexity of the MAML-based predictor can be approximated to $\mathcal{O}(N_{\text{epoch}}T_\text{S}N_\text{mt}M^2)$. After the meta-training stage, the complexity of the meta-adaptation stage becomes $\mathcal{O}(T_{\text{ad}}N_\text{ad}M^2)$, which is much lower than that of the meta-training stage since $T_\text{S}N_{\text{mt}} \gg T_\text{ad}N_\text{ad}$. In practice, the BS can perform the meta-training stage in advance, and only the meta-adaptation stage is needed online to predict the channels of certain UEs. Thus, the BS can achieve fast adaptive channel prediction using the MAML algorithm.

The DIP-based denoising process exploits the CNN structure as in Fig \ref{figure4}. Since the complexity of the CNN is dominated by the convolutions, we only consider the complexity of the convolution operations for the DIP-based denoising process. In the $i$-th convolutional layer, a group of $M_{i}$ filters of the $1 \times 1$ convolution are applied to $M_{i}$ feature maps of the dimension $N_{i} \times 1$. The complexity of the DIP-based denoising process with the number of iterations $N_\text{iter}$ is given by \cite{Taghavi19}
\begin{align}
C_\text{DIP}&=\mathcal{O}\left(N_\text{iter}\left(\sum_{i=1}^{L_d}N_{i}M_{i}^2+N_{L_d+1}M_{L_d+1}M\right)\right)\notag\\
&\stackrel{(a)}{=}\mathcal{O}\left(N_\text{iter}\left(\sum_{i=1}^{L_d}2^{i-1}N_1M_{i}^2+2^{L_d-1}N_1M_{L_d+1}M\right)\right)\notag\\
&\stackrel{(b)}{=}\mathcal{O}\left(N_\text{iter}\left(\left(2^{L_d}-1\right)N_1N_f^2+2^{{L_d}-1}N_1N_fM\right)\right)\notag\\
&\stackrel{(c)}{=}\mathcal{O}\left(N_\text{iter}N_t\left(2N_f^2+N_fM\right)\right)\notag\\
&=\mathcal{O}\left(N_\text{iter}N_tN_f(2N_f+M)\right),
\end{align} where $(a)$ comes from $N_i=2^{i-1}N_1$ for $1\leq i \leq L_d$ and $N_{L_d+1}=N_{L_d}$, $(b)$ is from the assumption of using the same number of filters $N_f$ in each layer, i.e., $M_i=N_f$ for all $i$, and $(c)$ is derived by $N_t\triangleq2^{L_d-1}N_1$ and $\left(2^{L_d}-1\right)N_1N_f \approx 2^{L_d}N_1N_f$. The complexity of the DIP-based denoising process can be approximated as $\mathcal{O}(N_{\text{iter}}N_tN_fM)$ for large $M$.
The computational complexity of the proposed MAML channel prediction algorithm and the DIP-based denoising process is summarized in Table I.

\subsection{Numerical Results}
The ML algorithms are implemented by a TensorFlow 2.0 and a NVIDIA Quadro RTX 8000 GPU for the numerical simulations. We perform Monte-Carlo simulations to verify the proposed channel prediction algorithm. In this paper, we employ the normalized mean-squared error (NMSE)
\begin{align}
\text{NMSE}=\mathbb{E}\left[{\left\|\hat{{\bh}}_{n+1,k_\text{T}}-{\bh}_{n+1,k_\text{T}}\right\|}^2/\norm{{\bh}_{n+1,k_\text{T}}}^2\right],
\end{align} for the performance metric. Also, we use the achievable sum-rate as the performance metric. To reduce the inter-user interference, the zero-forcing (ZF) combiner is adopted to
\begin{align}
\bar{\bF}_{n}^\mathrm{T}=\left(\hat{{\bH}}_{n}^\mathrm{H}\hat{{\bH}}_{n}\right)^{-1}\hat{{\bH}}_{n}^\mathrm{H},
\end{align}
with the predicted channel matrix $\hat{{\bH}}_{n}=\begin{bmatrix}\hat{\bh}_{n,1} \cdots \hat{\bh}_{n,K_t}\end{bmatrix}$. Note that $K_t$ is the number of UEs in the target task.
 We obtain the unit-norm combiner $\bff_{n,k_\text{T}}=\bar{\bff}_{n,k_\text{T}}/{\norm{\bar{\bff}_{n,k_\text{T}}}}$, where $\bar{\bff}_{n,k_\text{T}}$ represents the $k_\text{T}$-th columns of $\tilde{\bF}_{n}$. For the $k_\text{T}$-th UE, the achievable rate based on the receive combiner $\bff_{n,k_\text{T}}$ can be expressed as
\begin{align}
R_{k_\text{T}}=\log_2{\left(1+\frac{\rho|\bff_{n,k_\text{T}}^\mathrm{T}\bh_{n,k_\text{T}}|^2}{\rho\sum_{i \neq k_\text{T}}|\bff_{n,k_\text{T}}^\mathrm{T}\bh_{n,i}|^2+1}\right)}. \label{achievable rate}
\end{align}The achievable sum-rate is defined as
\begin{align}
R=\sum_{k_\text{T}=1}^{K_t} R_{k_\text{T}}. \label{achievable sum-rate}
\end{align}

\begin{table}[t]
	\renewcommand{\arraystretch}{1.3}
	\captionsetup{justification=centering}
	\captionsetup{labelsep=newline}
	\caption{System parameters}
	\centering
	\label{table2}
	
	\begin{tabular}{l  l}
		\toprule
		Parameter & Value \\
		\midrule
		Environment & UMi\\
		Carrier frequency & 2.3 GHz\\
		UE mobility & {3} km/h\\
		Time slot duration  & {40 ms}\\
		Number of BS antenna & 64\\
		Number of source tasks per UE & 1024 \\
	    Complexity order & 3\\
		Number of epochs & 20 \\
		Batch size & 64 \\
		Number of sample pairs in support set & 10\\
		Number of sample pairs in query set & 10\\
		Inner-task learning rate & $10^{-1}$ \\
		Outer-task learning rate & $10^{-5}$ \\
		\bottomrule
	\end{tabular}
\end{table}

We assume the spatial channel model (SCM) urban micro (UMi) scenario in \cite{SCM} with carrier frequency $f_c=2.3$ GHz, UE mobility $v=3$ km/h, and time slot duration $T_d=40$ ms. We use the MLP as the NN structure for the MAML algorithm based on the $L=4$ hidden-layers with $512$ nodes. For the DIP architecture, we adopt the CNN with the number of iterations $N_\text{iter}=2000$ and $L_d=4$ hidden-layers including $M_i=64$ for all $i$. We also set the number of BS antennas $M=64$, the number of source tasks per UE $T_u=1024$, the complexity order $n_o=3$, the number of epochs $N_\text{epoch}=20$, and the batch size $V=64$. The number of sample pairs of the support set is $N_s=10$, and the number of sample pairs of the query set is $N_q=10$. The inner-task learning rate is set to $\alpha=10^{-1}$, and the outer-task learning rate is set to $\beta=10^{-5}$. The system parameters are summarized in Table II.

\begin{figure}[t]
	\centering
	\includegraphics[width=9 cm]{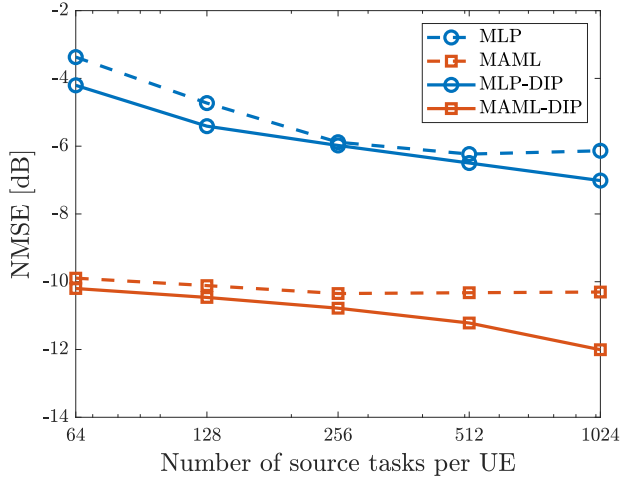}
	\caption{NMSE vs. number of source tasks per UE with $T_{\text{ad}}=10$, $N_{\text{ad}}=20$, and $\text{SNR}=20$ dB.}
	\label{figure5}
\end{figure}

\begin{figure}[t]
	\centering
	\includegraphics[width=9 cm]{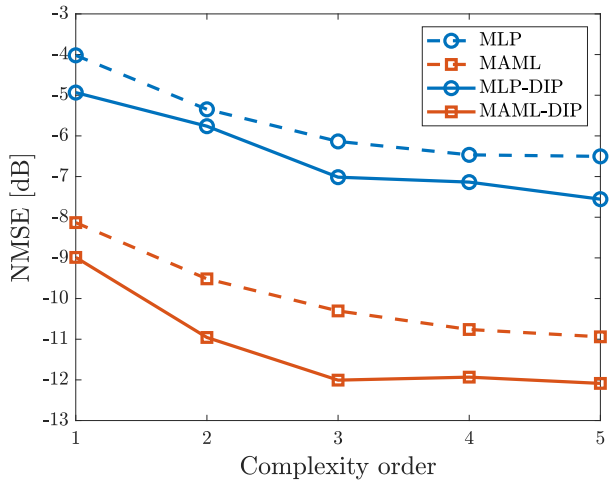}
	\caption{NMSE vs. complexity order with $T_{\text{ad}}=10$, $N_{\text{ad}}=20$, and $\text{SNR}=20$ dB.}
	\label{figure6}
\end{figure}

\begin{figure}[t]
	\centering
	\includegraphics[width=9 cm]{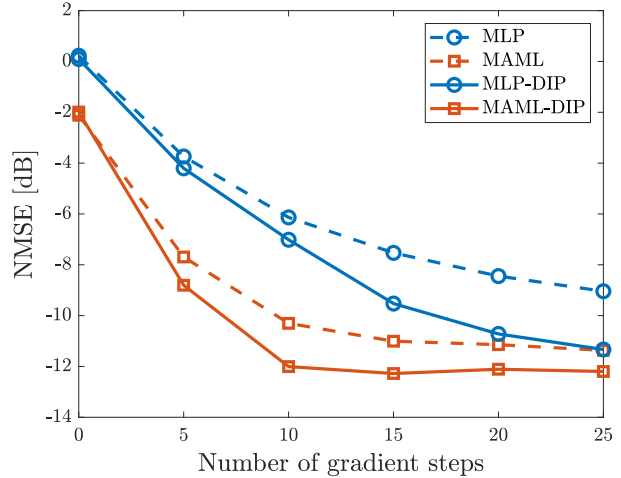}
	\caption{NMSE vs. number of gradient steps with $N_{\text{ad}}=20$ and $\text{SNR}=20$ dB.}
	\label{figure7}
\end{figure}

In the simulation, we compare the following predictors:
\begin{itemize}
	\item \textbf{MLP}: first optimized with the source dataset and re-trained with a few samples from the target dataset without the denoising process. This serves as a baseline of our predictor.
	\item \textbf{MAML}: proposed MAML-based prediction without the denoising process.
	\item \textbf{MLP-DIP}: MLP with the denoised LS channel estimate based on the DIP.
	\item \textbf{MAML-DIP}: proposed MAML-based prediction with the denoised LS channel estimate based on the DIP.
\end{itemize}

\noindent {Note that all these four methods rely on the same source and target tasks, and the main difference between the MLP-based and the MAML-based predictors is the structure of NN.

We consider the LS channel estimates for a total of 8 UEs with the number of UEs in the source task $K_s=4$ and the number of UEs in the target task $K_t=4$.} The BS trains each network with the first 4 UE LS channel estimates from the source dataset $\mathbb{D}_\text{S}$ and test with the remaining 4 UE LS channel estimates from the target dataset $\mathbb{D}_\text{T}$ to obtain the average NMSE and sum-rate with $N_\text{te}=100$. 

To verify the effect of the number of source tasks per UE $T_u$, Fig. \ref{figure5} shows the NMSEs of the MLP channel prediction and the MAML channel prediction as a function of $T_u$. In this simulation, we assume that the number of gradient steps $T_{\text{ad}}=10$, the number of adaptation samples $N_{\text{ad}}=20$, and $\text{SNR}=20$ dB. The NMSEs of both channel predictions without the DIP decrease as the number of source tasks per UE increases, but eventually saturate. The denoising process is able to break this saturation effect on both methods while the gain of denoising is larger for the MAML channel prediction. We set the number of source tasks per UE at $T_u=1024$ for the following simulations.

In Fig. \ref{figure6}, we compare the MAML channel prediction to the MLP channel prediction in terms of NMSE according to the complexity order with $T_\text{ad}=10$, $N_\text{ad}=20$, and $\text{SNR}=20$ dB. The NMSEs of both channel predictions decrease as the complexity order increases until $n_o=3$, but the gain becomes marginal after. Therefore, we set $n_o=3$ to balance the accuracy and complexity of channel predictions in the following simulations. Note that the complexity order needs to be larger to achieve the same accuracy when the UE mobility increases \cite{Kim21}.

\begin{figure}[t]
	\centering
	\includegraphics[width=9 cm]{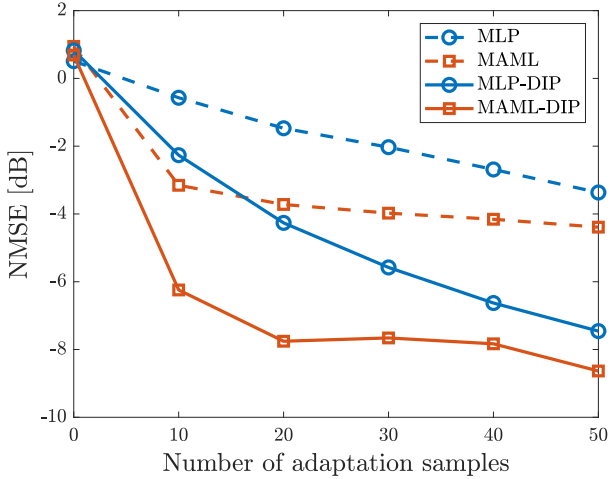}%
	\caption{NMSE vs. number of adaptation samples with $\text{SNR}=0$ dB.}
	\label{figure8}
\end{figure}

\begin{figure}[t]
	\centering
	\includegraphics[width=9 cm]{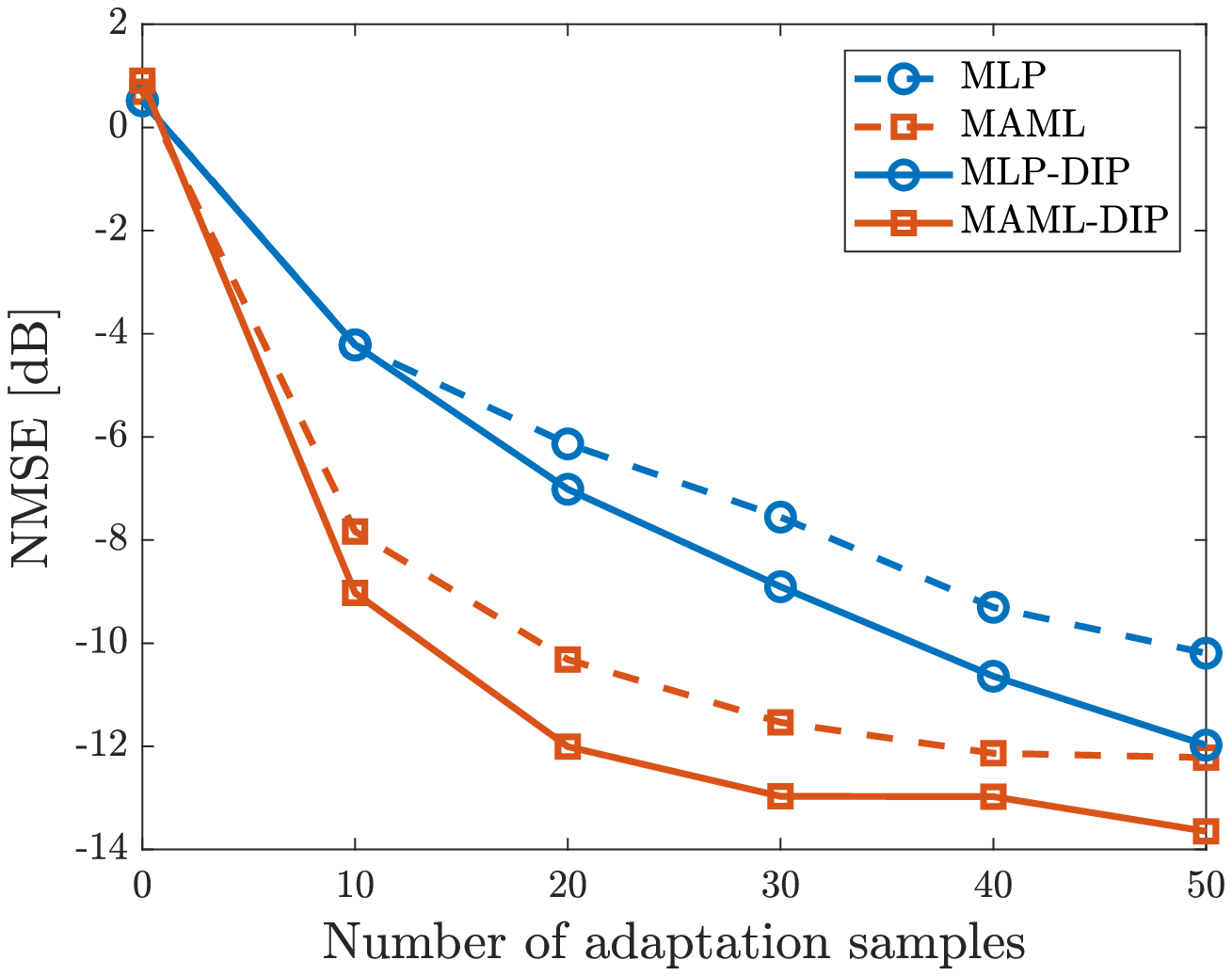}%
	\caption{NMSE vs. number of adaptation samples with $\text{SNR}=20$ dB.}
	\label{figure9}
\end{figure}

\begin{figure}[t]
	\centering
	\includegraphics[width=9 cm]{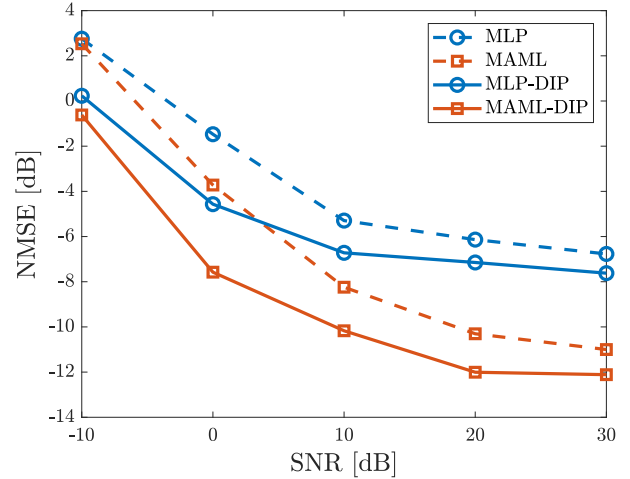}%
	\caption{NMSE vs. SNR with $N_\text{ad}=20$.}
	\label{figure10}
\end{figure}

\begin{figure}[t]
	\centering
	\includegraphics[width=9 cm]{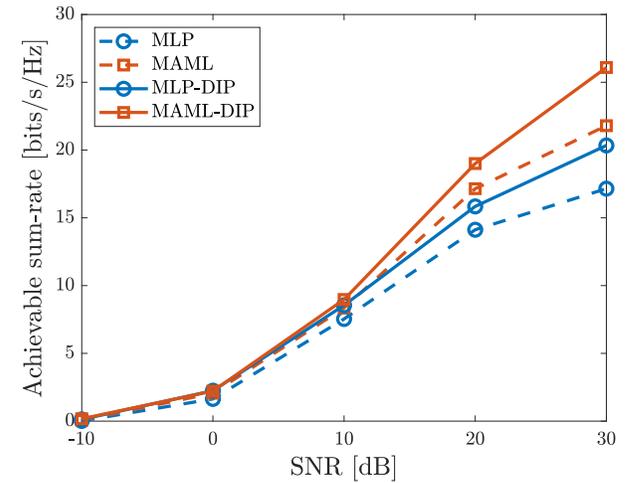}
	\caption{Achievable sum-rate vs. SNR with $N_\text{ad}=20$.}
	\label{figure11}
\end{figure}

Fig. \ref{figure7} shows the NMSEs of the MLP channel prediction and the MAML channel prediction according to the number of gradient steps with $N_\text{ad}=20$ and $\text{SNR}=20$ dB. The figure shows that the proposed MAML channel prediction gives a moderate gain compared to the MLP channel prediction regardless of the number of gradient steps. In addition, the NMSEs of the MAML channel predictions almost converge when the number of gradient steps reaches 10 while the MLP channel prediction requires to have more gradient steps to converge. In the following simulations, we set $T_\text{Ad}=10$. 

In Figs. \ref{figure8} and \ref{figure9}, the NMSEs of both the MLP and MAML channel predictions are compared according to the number of adaptation samples with different SNR values. The figures clearly show that the MAML channel prediction outperforms the MLP channel prediction. Also, the MAML channel prediction has moderate accuracy with a small number of adaptation samples, e.g., $20$ and $30$ adaptation samples for $0$ dB and $20$ dB SNR values, respectively. Moreover, the DIP-based denoising process gives additional $4$ dB gain when the SNR is $0$ dB and $2$ dB gain when the SNR is $20$ dB.

Fig. \ref{figure10} plots the NMSEs of the MLP and MAML channel predictions according to the SNR with $N_\text{ad}=20$. The NMSEs of all cases saturate as the SNR increases but the gap between the MAML and MLP channel predictions remains. Although the effect of noise will eventually become negligible as the SNR increases, the gain of the DIP exists even for quite large SNR values. 

Fig. \ref{figure11} depicts the achievable sum-rates of the MLP and MAML channel predictions according to the SNR with $N_\text{ad}=20$. Because of better prediction quality, the MAML channel prediction can achieve much higher achievable sum-rate than the MLP channel prediction when the SNR is large enough. The DIP-based denoising process further boosts the achievable sum-rate, which shows the importance of data preprocessing before training the NN. Also, the DIP-based denoising process provides an increased achievable sum-rate across all SNRs.

\section{Conclusion}\label{Section:conclusion}
In this paper, we proposed a fast adaptive channel predictor for massive MIMO systems using the MAML algorithm, which is the popular meta-learning technique. The proposed MAML channel prediction extracts the key characteristics of time-varying channels and exploits these features to adaptively predict channels in new environments. Also, the DIP-based denoising process applied to the training samples further improves the prediction performance by reducing the noise effect. Numerical results showed that the MAML channel prediction provides improvements in the complexity, accuracy, and achievable sum-rate even with only a few adaptation samples. These improvements make the MAML channel prediction highly practical.

\bibliographystyle{IEEEtran}
\bibliography{IEEEtwc}

\begin{thebibliography}{10}
\providecommand{\url}[1]{#1}
\csname url@samestyle\endcsname
\providecommand{\newblock}{\relax}
\providecommand{\bibinfo}[2]{#2}
\providecommand{\BIBentrySTDinterwordspacing}{\spaceskip=0pt\relax}
\providecommand{\BIBentryALTinterwordstretchfactor}{4}
\providecommand{\BIBentryALTinterwordspacing}{\spaceskip=\fontdimen2\font plus
\BIBentryALTinterwordstretchfactor\fontdimen3\font minus
  \fontdimen4\font\relax}
\providecommand{\BIBforeignlanguage}[2]{{%
\expandafter\ifx\csname l@#1\endcsname\relax
\typeout{** WARNING: IEEEtran.bst: No hyphenation pattern has been}%
\typeout{** loaded for the language `#1'. Using the pattern for}%
\typeout{** the default language instead.}%
\else
\language=\csname l@#1\endcsname
\fi
#2}}
\providecommand{\BIBdecl}{\relax}
\BIBdecl

\bibitem{Marzetta06}
T.~L. Marzetta, ``Noncooperative {c}ellular {w}ireless with {u}nlimited
  {n}umbers of {b}ase {s}tation {a}ntennas,'' \emph{IEEE Transactions on
  Wireless Communications}, vol.~9, no.~11, pp. 3590--3600, Nov. 2010.

\bibitem{Papa17}
A.~K. {Papazafeiropoulos}, ``Impact of general channel aging conditions on the
  downlink performance of massive {MIMO},'' \emph{IEEE Transactions on
  Vehicular Technology}, vol.~66, no.~2, pp. 1428--1442, Feb. 2017.

\bibitem{Truong13}
K.~T. {Truong} and R.~W. {Heath}, ``Effects of channel aging in massive {MIMO}
  systems,'' \emph{Journal of Communications and Networks}, vol.~15, no.~4, pp.
  338--351, Aug. 2013.

\bibitem{Choi14}
J.~Choi, D.~J. Love, and P.~Bidigare, ``Downlink training techniques for {FDD}
  massive {MIMO} systems: Open-loop and closed-loop training with memory,''
  \emph{IEEE Journal of Selected Topics in Signal Processing}, vol.~8, no.~5,
  pp. 802--814, Oct. 2014.

\bibitem{Kong15}
C.~{Kong}, C.~{Zhong}, A.~K. {Papazafeiropoulos}, M.~{Matthaiou}, and
  Z.~{Zhang}, ``Sum-rate and power scaling of massive {MIMO} systems with
  channel aging,'' \emph{IEEE Transactions on Communications}, vol.~63, no.~12,
  pp. 4879--4893, Dec. 2015.

\bibitem{Larew19}
S.~G. Larew and D.~J. Love, ``Adaptive beam tracking with the unscented
  {K}alman filter for millimeter wave communication,'' \emph{IEEE Signal
  Processing Letters}, vol.~26, no.~11, pp. 1658--1662, Nov. 2019.

\bibitem{Chou21}
T.-H. Chou, N.~Michelusi, D.~J. Love, and J.~V. Krogmeier, ``Fast
  position-aided {MIMO} beam training via noisy tensor completion,'' \emph{IEEE
  Journal of Selected Topics in Signal Processing}, vol.~15, no.~3, pp.
  774--788, Apr. 2021.

\bibitem{AIML}
\emph{Study on {A}rtificial {I}ntelligence ({AI})/{M}achine Learning ({ML}) for
  {NR} {A}ir {I}nterface}, 3GPP TR 38.843 Std., Dec. 2021.

\bibitem{Dong19}
P.~{Dong}, H.~{Zhang}, G.~Y. {Li}, N.~{NaderiAlizadeh}, and I.~S. {Gaspar},
  ``Deep {CNN} for wideband mmwave massive {MIMO} channel estimation using
  frequency correlation,'' in \emph{2019 IEEE International Conference on
  Acoustics, Speech and Signal Processing (ICASSP)}, May 2019, pp. 4529--4533.

\bibitem{Jiang20}
W.~Jiang, M.~Strufe, and H.~Dieter~Schotten, ``Long-range {MIMO} channel
  prediction using recurrent neural networks,'' in \emph{2020 IEEE 17th Annual
  Consumer Communications Networking Conference (CCNC)}, 2020, pp. 1--6.

\bibitem{Yuan20}
J.~Yuan, H.~Q. Ngo, and M.~Matthaiou, ``Machine learning-based channel
  prediction in massive {MIMO} with channel aging,'' \emph{IEEE Transactions on
  Wireless Communications}, vol.~19, no.~5, pp. 2960--2973, May 2020.

\bibitem{Bogale20}
T.~E. Bogale, X.~Wang, and L.~B. Le, ``Adaptive channel prediction, beamforming
  and scheduling design for {5G V2I} network: Analytical and machine learning
  approaches,'' \emph{IEEE Transactions on Vehicular Technology}, vol.~69,
  no.~5, pp. 5055--5067, May 2020.

\bibitem{Kim21}
H.~Kim, S.~Kim, H.~Lee, C.~Jang, Y.~Choi, and J.~Choi, ``Massive {MIMO} channel
  prediction: Kalman filtering vs. machine learning,'' \emph{IEEE Transactions
  on Communications}, vol.~69, no.~1, pp. 518--528, Jan. 2021.

\bibitem{Wu21}
C.~Wu, X.~Yi, Y.~Zhu, W.~Wang, L.~You, and X.~Gao, ``Channel prediction in
  high-mobility massive {MIMO}: From spatio-temporal autoregression to deep
  learning,'' \emph{IEEE Journal on Selected Areas in Communications}, vol.~39,
  no.~7, pp. 1915--1930, Jul. 2021.

\bibitem{Andrychowicz16}
M.~Andrychowicz, M.~Denil, S.~G\'{o}mez, M.~W. Hoffman, D.~Pfau, T.~Schaul,
  B.~Shillingford, and N.~de~Freitas, ``Learning to learn by gradient descent
  by gradient descent,'' in \emph{Advances in Neural Information Processing
  Systems}, vol.~29, 2016.

\bibitem{Ravi17}
S.~Ravi and H.~Larochelle, ``Optimization as a model for few-shot learning,''
  in \emph{International Conference on Learning Representations}, 2017.

\bibitem{HWu19}
H.~Wu, Z.~Zhang, C.~Jiao, C.~Li, and T.~Q.~S. Quek, ``Learn to sense: A
  meta-learning-based sensing and fusion framework for wireless sensor
  networks,'' \emph{IEEE Internet of Things Journal}, vol.~6, no.~5, pp.
  8215--8227, Oct. 2019.

\bibitem{Mao19}
H.~Mao, H.~Lu, Y.~Lu, and D.~Zhu, ``Roemnet: Robust meta learning based channel
  estimation in {OFDM} systems,'' in \emph{ICC 2019 - 2019 IEEE International
  Conference on Communications (ICC)}, 2019, pp. 1--6.

\bibitem{Zhang21}
J.~Zhang, Y.~He, Y.-W. Li, C.-K. Wen, and S.~Jin, ``Meta learning-based {MIMO}
  detectors: Design, simulation, and experimental test,'' \emph{IEEE
  Transactions on Wireless Communications}, vol.~20, no.~2, pp. 1122--1137,
  Feb. 2021.

\bibitem{Park21}
S.~Park, H.~Jang, O.~Simeone, and J.~Kang, ``Learning to demodulate from few
  pilots via offline and online meta-learning,'' \emph{IEEE Transactions on
  Signal Processing}, vol.~69, pp. 226--239, Jan. 2021.

\bibitem{Yuan21}
Y.~Yuan, G.~Zheng, K.-K. Wong, B.~Ottersten, and Z.-Q. Luo, ``Transfer learning
  and meta learning-based fast downlink beamforming adaptation,'' \emph{IEEE
  Transactions on Wireless Communications}, vol.~20, no.~3, pp. 1742--1755,
  Mar. 2021.

\bibitem{Xia21}
J.~Xia and D.~Gunduz, ``Meta-learning based beamforming design for {MISO}
  downlink,'' in \emph{2021 IEEE International Symposium on Information Theory
  (ISIT)}, Jul. 2021, pp. 2954--2959.

\bibitem{Long21}
Y.~Long and S.~Murphy, ``Few-shot learning based hybrid beamforming under
  birth-death process of scattering paths,'' \emph{IEEE Communications
  Letters}, vol.~25, no.~5, pp. 1687--1691, May 2021.

\bibitem{Zhang22}
J.~Zhang, Y.~Yuan, G.~Zheng, I.~Krikidis, and K.-K. Wong, ``Embedding
  model-based fast meta learning for downlink beamforming adaptation,''
  \emph{IEEE Transactions on Wireless Communications}, vol.~21, no.~1, pp.
  149--162, Jan. 2022.

\bibitem{Yang20}
Y.~Yang, F.~Gao, Z.~Zhong, B.~Ai, and A.~Alkhateeb, ``Deep transfer
  learning-based downlink channel prediction for {FDD} massive {MIMO}
  systems,'' \emph{IEEE Transactions on Communications}, vol.~68, no.~12, pp.
  7485--7497, Dec. 2020.

\bibitem{Zeng21}
J.~Zeng, J.~Sun, G.~Gui, B.~Adebisi, T.~Ohtsuki, H.~Gacanin, and H.~Sari,
  ``Downlink {CSI} feedback algorithm with deep transfer learning for {FDD}
  massive {MIMO} systems,'' \emph{IEEE Transactions on Cognitive Communications
  and Networking}, vol.~7, no.~4, pp. 1253--1265, Dec. 2021.

\bibitem{Finn17}
C.~Finn, P.~Abbeel, and S.~Levine, ``Model-agnostic meta-learning for fast
  adaptation of deep networks,'' in \emph{Proceedings of the 34th International
  Conference on Machine Learning}, vol.~70, Aug. 2017, pp. 1126--1135.

\bibitem{Kim20}
H.~Kim, S.~Kim, H.~Lee, and J.~Choi, ``Massive {MIMO} channel prediction:
  Machine learning versus {K}alman filtering,'' in \emph{2020 IEEE Globecom
  Workshops (GC Wkshps)}, Dec. 2020, pp. 1--6.

\bibitem{Jin19}
Y.~Jin, J.~Zhang, S.~Jin, and B.~Ai, ``Channel estimation for cell-free mmwave
  massive {MIMO} through deep learning,'' \emph{IEEE Transactions on Vehicular
  Technology}, vol.~68, no.~10, pp. 10\,325--10\,329, Oct. 2019.

\bibitem{Soltani19}
M.~Soltani, V.~Pourahmadi, A.~Mirzaei, and H.~Sheikhzadeh, ``Deep
  learning-based channel estimation,'' \emph{IEEE Communications Letters},
  vol.~23, no.~4, pp. 652--655, Apr. 2019.

\bibitem{Balevi20}
E.~Balevi, A.~Doshi, and J.~G. Andrews, ``Massive {MIMO} channel estimation
  with an untrained deep neural network,'' \emph{IEEE Transactions on Wireless
  Communications}, vol.~19, no.~3, pp. 2079--2090, Mar. 2020.

\bibitem{Zhang17}
K.~Zhang, W.~Zuo, Y.~Chen, D.~Meng, and L.~Zhang, ``Beyond a {G}aussian
  denoiser: Residual learning of deep {CNN} for image denoising,'' \emph{IEEE
  Transactions on Image Processing}, vol.~26, no.~7, pp. 3142--3155, Jul. 2017.

\bibitem{He18}
H.~He, C.-K. Wen, S.~Jin, and G.~Y. Li, ``Deep learning-based channel
  estimation for beamspace mmwave massive {MIMO} systems,'' \emph{IEEE Wireless
  Communications Letters}, vol.~7, no.~5, pp. 852--855, Oct. 2018.

\bibitem{Zhang20}
Y.~Zhang, Y.~Mu, Y.~Liu, T.~Zhang, and Y.~Qian, ``Deep learning-based beamspace
  channel estimation in mmwave massive {MIMO} systems,'' \emph{IEEE Wireless
  Communications Letters}, vol.~9, no.~12, pp. 2212--2215, Dec. 2020.

\bibitem{Ye20}
H.~Ye, F.~Gao, J.~Qian, H.~Wang, and G.~Y. Li, ``Deep learning-based denoise
  network for {CSI} feedback in {FDD} massive {MIMO} systems,'' \emph{IEEE
  Communications Letters}, vol.~24, no.~8, pp. 1742--1746, Aug. 2020.

\bibitem{Ulyanov18}
D.~Ulyanov, A.~Vedaldi, and V.~Lempitsky, ``Deep image prior,'' in
  \emph{Proceedings of the IEEE Conference on Computer Vision and Pattern
  Recognition (CVPR)}, Jun. 2018.

\bibitem{Hung10}
K.-C. Hung and D.~W. Lin, ``Pilot-based {LMMSE} channel estimation for {OFDM}
  systems with power–delay profile approximation,'' \emph{IEEE Transactions
  on Vehicular Technology}, vol.~59, no.~1, pp. 150--159, Jan. 2010.

\bibitem{Timothy21}
T.~M. Hospedales, A.~Antoniou, P.~Micaelli, and A.~J. Storkey, ``Meta-learning
  in neural networks: A survey,'' \emph{IEEE Transactions on Pattern Analysis
  and Machine Intelligence}, pp. 1--1, 2021.

\bibitem{Goyal17}
\BIBentryALTinterwordspacing
P.~Goyal, P.~Dollár, R.~Girshick, P.~Noordhuis, L.~Wesolowski, A.~Kyrola,
  A.~Tulloch, Y.~Jia, and K.~He, ``Accurate, large minibatch {SGD}: Training
  imagenet in 1 hour,'' 2017. [Online]. Available:
  \url{https://arxiv.org/abs/1706.02677}
\BIBentrySTDinterwordspacing

\bibitem{Kingma15}
\BIBentryALTinterwordspacing
D.~P. Kingma and J.~Ba, ``Adam: A method for stochastic optimization,'' in
  \emph{ICLR (Poster)}, 2015. [Online]. Available:
  \url{http://arxiv.org/abs/1412.6980}
\BIBentrySTDinterwordspacing

\bibitem{Arridge19}
S.~Arridge, P.~Maass, O.~Öktem, and C.-B. Schönlieb, ``Solving inverse
  problems using data-driven models,'' \emph{Acta Numerica}, vol.~28, p.
  1–174, 2019.

\bibitem{He21}
K.~He, L.~He, L.~Fan, Y.~Deng, G.~K. Karagiannidis, and A.~Nallanathan,
  ``Learning-based signal detection for {MIMO} systems with unknown noise
  statistics,'' \emph{IEEE Transactions on Communications}, vol.~69, no.~5, pp.
  3025--3038, May 2021.

\bibitem{Hunger05}
R.~Hunger, \emph{Floating point operations in matrix-vector calculus}.\hskip
  1em plus 0.5em minus 0.4em\relax Munich University of Technology, Inst. for
  Circuit Theory and Signal, 2005.

\bibitem{Mizutani01}
E.~{Mizutani} and S.~E. {Dreyfus}, ``On complexity analysis of supervised
  {MLP}-learning for algorithmic comparisons,'' in \emph{IJCNN'01.
  International Joint Conference on Neural Networks. Proceedings (Cat.
  No.01CH37222)}, vol.~1, Jul. 2001, pp. 347--352.

\bibitem{Taghavi19}
M.~Taghavi and M.~Shoaran, ``Hardware complexity analysis of deep neural
  networks and decision tree ensembles for real-time neural data
  classification,'' in \emph{2019 9th International IEEE/EMBS Conference on
  Neural Engineering (NER)}, Mar. 2019, pp. 407--410.

\bibitem{SCM}
\emph{Study on 3{D} channel model for {LTE}}, 3GPP TR 36.873 V12.7.0 Std., Jan.
  2018.

\end{thebibliography}


\end{document}